\pdfoutput=1

\newcommand\be{\begin{equation}}
\newcommand\ee{\end{equation}}
\newcommand\bd{\begin{displaymath}} 
\newcommand\ed{\end{displaymath}}

\documentclass{emulateapj}
\usepackage{apjfonts,graphicx,lscape,here}
\usepackage{common} 
\begin{document}
\def\gae{\mathrel{\hbox{\rlap{\hbox{\lower2pt\hbox{$\sim$}}}\hbox{\raise2pt\hbox{$>$}}}}}
\def\lae{\mathrel{\hbox{\rlap{\hbox{\lower2pt\hbox{$\sim$}}}\hbox{\raise2pt\hbox{$<$}}}}}
\def\msunyr{{\rm\, M_\odot\, yr^{-1}}}

\title{A Ten Billion Solar Mass Outflow of Molecular Gas Launched by Radio Bubbles in the Abell 1835 Brightest Cluster Galaxy}

\author{B. R. McNamara$^{1,2,3}$ H.R. Russell$^1$, P. E. J. Nulsen$^3$
A. C. Edge$^4$, N. W. Murray$^7$, R. A. Main$^1$, A. N. Vantyghem$^1$, F. Combes$^5$, A. C. Fabian$^6$, P. Salome$^5$, C.C. Kirkpatrick$^1$ , S. A. Baum$^8$, J. N. Bregman$^9$, M. Donahue$^{10}$, E. Egami$^{11}$, S. Hamer$^5$, C. P. O'Dea$^8$ , J.B.R. Oonk$^{12}$, G. Tremblay$^{13}$, G.M. Voit$^{10}$}

\affil{$^1$University of Waterloo, Department of Physics \& Astronomy, Waterloo, Canada $^2$Perimeter Institute for Theoretical Physics, Waterloo, Canada $^3$Harvard-Smithsonian Center for Astrophysics $^4$ Department of Physics, Durham University, Durham, DH1 3LE,  UK 
$^5$ L'Observatoire de Paris, 61 Av. de L'Observatoire, F-75 014 Paris, France 
$^6$	Institute of Astronomy,  Madingley Road, Cambridge, CB3 0HA, UK 
$^7$ Canadian Institute for Theoretical Astrophysics, University of Toronto, 60 St. George St., Toronto, M5S 3H8, Ontario, Canada
$^8$ School of Physics \& Astronomy, Rochester Institute of Technology, Rochester, NY 14623 USA
$^9$ Department of Astronomy, University of Michigan,  500 Church St. Ann Arbor, MI 48109 USA
$^{10}$Department of Physics \& Astronomy, Michigan State University, 567 Wilson Rd., East Lansing, MI 48824 USA
$^{11}$ Steward Observatory, University of Arizona, 933 N. Cherry Avenue, Tucson, AZ 85721 USA
$^{12}$ Sterrewacht Leiden, Universiteit Leiden, P.O. Box 9513, NL-2300 RA Leiden, Netherlands 
$^{13}$European Southern Observatory, Karl-Schwarzschild-Strasse 2, 85748 Garching, Germany 
}


\begin{abstract}
We report ALMA Early Science observations of the Abell 1835 brightest cluster galaxy (BCG) in the CO (3-2) and
CO (1-0)  emission lines.  We detect $5\times 10^{10}~\rm M_\odot$ of molecular gas within 10 kpc of the
BCG.   Its velocity width of $\sim 130 ~\rm km~s^{-1}$ FWHM  is too narrow to be supported by dynamical pressure.  The gas may 
instead be supported in a rotating, turbulent disk oriented nearly face-on.  The disk is forming stars at a rate of $100-180~\rm M_\odot ~yr^{-1}$.
Roughly $10^{10}~\rm M_\odot$ of molecular gas is projected $3-10 ~\rm kpc$ to the
north-west and to the east of the nucleus with line of sight velocities lying between $-250 ~\rm km~s^{-1}$ to $+480 ~\rm km~s^{-1}$  with respect to the systemic velocity.   Although inflow cannot be ruled out, the rising velocity gradient with radius is consistent with a broad,  bipolar outflow driven by radio jets or buoyantly rising X-ray cavities.  The molecular outflow may be associated with 
an outflow of hot gas in Abell 1835 seen on larger scales.  Molecular gas is
flowing out of the BCG at a rate of approximately $200~\rm M_\odot ~yr^{-1}$, which is comparable to its star formation rate.  
How radio bubbles lift  dense molecular gas in their updrafts,  how much gas will be lost to the BCG,  and how much will return to fuel future star formation and AGN activity are poorly understood.  Our results imply that radio-mechanical (radio mode) feedback not only heats hot atmospheres surrounding elliptical galaxies and BCGs, it is able to sweep higher density molecular gas away from their centers.  

\end{abstract}


\section{Introduction}
Brightest cluster galaxies (BCGs) are the largest and most luminous galaxies in the universe.  Like normal elliptical galaxies, their stellar populations are usually old and dormant.  BCGs residing in  cooling flow clusters are exceptional (Fabian 1994).  Many form stars at rates of several to several tens of solar masses per year
(O'Dea et al. 2008), fueled by unusually large reservoirs of cold molecular clouds (Edge et al. 2001, Salome \& Combes 2003). Extreme objects, such as the Phoenix  and the Abell 1835 BCGs, are forming stars at rates upward of $100~\rm M_\odot ~yr^{-1}$ (McDonald et al. 2012, McNamara et al. 2006, hereafter M06).   

Galaxies may be rejuvenated by collisions.  However, wet mergers are unlikely to occur at the centers of rich galaxy clusters.   Molecular clouds and young stars forming in BCGs are apparently fueled instead
by gas cooling from hot atmospheres.  Bright nebular emission and young stars are observed preferentially when the central cooling time of a cluster atmosphere falls below $\sim 1$ Gyr (Heckman 1981, Hu et al. 1985).  
High resolution X-ray imaging has since revealed that nebular emission and star formation appear at a sharp
threshold or transition as the central cooling time falls below $\sim 5 \times 10^8~\rm yr$  (Rafferty et al. 2008, Cavagnolo et al. 2008).  This threshold is tied to cooling instabilities and thermal conduction in the hot atmosphere (Voit et al. 2008, Voit 2011, Gaspari et al. 2012, Guo \& Mathews 2013).

Despite strong indications that cold clouds are condensing out of hot atmospheres, only a few percent of the mass expected to cool actually does so (Peterson \& Fabian 2006). 
Feedback from active galactic nuclei (AGN) is almost certainly suppressing cooling below the levels expected in an unimpeded cooling flow (reviewed by McNamara \& Nulsen 2007, 2012, Fabian 2013).  So-called radio-mode or radio-mechanical feedback, is thought to operate primarily on
the hot, volume-filling atmosphere by increasing its entropy and, to some degree, by driving the most rapidly cooling
gas away from the center.  This mechanism regulates cooling and in turn governs star formation and the power output of the AGN.  

Little is known of the effect of radio-mechanical feedback on molecular gas.  Yet regulating
the rate of cold accretion onto AGN (Pizzolato \& Soker 2010, Gaspari et al. 2013) may be necessary to maintain the feedback loop and suppress star formation in elliptical galaxies and BCGs.   Radio jets located in powerful radio galaxies interact with ionized gas
(eg. Nesvadba et al. 2006), which is probably associated with molecular clouds (Wilman et al. 2006).  
Furthermore, blueshifted absorption lines of neutral atomic hydrogen have been observed toward several
radio galaxies (Morganti et al. 2005, 2013), indicating that radio jets couple to cold clouds and are able to drive them out at
high speed.  Flow rates are difficult to evaluate with absorption measurements, so their full impact on galaxy 
evolution is unclear.  This will change as sensitive, subarcsec imaging and emission line spectroscopy
of galaxies with the Atacama Large Millimeter Array (ALMA) become available.

Here we examine the effects of feedback on the molecular gas located near the nucleus of
the  Abell 1835 BCG.  The BCG contains upward of $\simeq 5 \times 10^{10}~\rm M_\odot$ of molecular
gas (Edge 2001) and star formation proceeding at a rate of between $100-180~\rm M_\odot ~yr^{-1}$ (M06). 
The AGN launched a pair of cavities into its hot atmosphere a few $10^7 \rm yr$ ago, each of which is 20 kpc in diameter and projected roughly 20 kpc  from the nucleus.   The AGN's radio synchrotron luminosity, $3.6\times 10^{41}~\rm erg~s^{-1}$, is dwarfed by its mechanical power, $ L_{\rm mec} \simeq 10^{45}~\rm erg s^{-1}$(M06), which is typical of radio AGN (Birzan et al. 2008).  Abell 1835 is an archetypal cooling flow regulated by radio-mode feedback.
The ALMA Early Science observations reported here and in a companion paper on Abell 1664 (Russell et al. 2013), 
 explore for the first time at high resolution, the relationships between molecular gas,  star formation, and radio AGN feedback.  At the emission line redshift z = 0.252 (Crawford et al. 1999), 1 arcsec = 3.9 kpc.

\section{Observations}

We obtained Early Science observations of the BCG with ALMA
at 92 GHz (band 3) and 276 GHz
(band 7).  At the cluster's redshift the bands are
sensitive to the carbon monoxide molecule's J = 1,0 and J = 3,2
rotational transitions, respectively. The exposures, totaling 60
minutes in band 3 and 60 minutes in band 7, were made between 2012
March 27 and 2012, April 24.  The extended array available for Cycle 0
included on average twenty 12 metre dishes, which provided a spatial
resolution of 0.5 arcsec in the CO (3-2) transition and 1.5 arcsec at
the CO (1-0) transition.  Baselines extended to
$\sim400\m$.  This combination yielded a sharp image of
the molecular gas near the nucleus at CO (3-2), and sensitivity on
larger spatial scales at CO (1-0).   A bandwidth of $1.875\GHz$ per 
spectral window and two spectral windows per
sideband provided a total frequency range of $\sim{7}\GHz$.  We used
a spectral resolution of $0.488\MHz$ per channel.  Channels
were binned together to improve the S/N ratio, yielding a final
resolution of $20\kmps$.  The quasar 3C\,279 was observed for bandpass
calibration and observations of Mars and Titan provided absolute flux
calibration.  Observations switched from Abell 1835 to the nearby
phase calibrator J1332+0200 every $\sim10$ minutes.


The observations were calibrated using the \textsc{casa} software
(version 3.3) following the detailed processing
scripts provided by the ALMA science support team.  The
continuum-subtracted images were reconstructed using the \textsc{casa}
task \textsc{clean} assuming Briggs weighting with a robustness
parameter of 0.5 and with a simple polygon mask applied to each
channel.  This provided a synthesized beam of
$1.7\arcsec\times1.3\arcsec$ at a PA of $-84.1^{\circ}$ at CO(1-0) and
$0.60\arcsec\times0.48\arcsec$ at a PA of $-80.0^{\circ}$ at CO(3-2).
The rms noise in the line free channels was $0.6~\rm mJy$ at CO(1-0) and
$1.6~\rm mJy$ at CO(3-2).  Images of the continuum emission were also
produced with \textsc{clean} by averaging channels free of any line
emission.  A central continuum source is detected in both bands at
position $14~01~02.083$, $+02~52~42.649$ with fluxes
$1.26 \pm 0.03 ~\rm mJy$ in band 3 and $0.7 \pm 0.1 \rm~ mJy$ in
band 7.  The mm-continuum source position coincides with the VLA radio
nucleus position (eg. Hogan et al. in prep).  The mm-continuum flux is
also consistent, within a factor of 2, with synchrotron emission from
a flat spectrum radio core\footnote{For the convention $f_{\nu}
  \propto \nu^{-\alpha}$} with $\alpha\propto0.84$ (Hogan et al. in
prep) suggesting this is the location of the low luminosity AGN.


\section{Analysis}
\subsection{Spectra}
The total CO (3-2) and CO (1-0) spectra are presented in Fig. 1.  CO emission is centered within $\sim 100~\rm km~s^{-1}$ of the nebular emission line redshift (Crawford et al. 1999).   Each  spectral profile was fitted with a single gaussian after the continuum was subtracted.  The emission integral at CO (1-0) is  $3.6 \pm 0.2~ \rm Jy ~km ~s^{-1}$.  The molecular gas mass was calculated as,

\begin{equation}
M_{\rm mol} = 1.05\times 10^4 ~X_{\rm CO}\left(\frac{1}{1+z}\right) \left (\frac{S_{\rm CO}\Delta v}{\rm Jy~km~s^{-1}}\right) \left(\frac{D_{\rm L}}{{\rm Mpc}}\right)^2~M_\odot.
\end{equation}

This expression yields a total molecular gas mass of $4.9 \pm 0.2 \times 10^{10} ~M_{\odot}$.  The conversion factor between CO and molecular gas, $\rm X_{CO} = 2 \times 10^{20}~ \rm cm^{-2} (K~ km ~s^{-1})^{-1}$, is the average Galactic value (reviewed by Bolatto et al. 2013).  The primary line at zero velocity has a gaussian profile full width at half maximum  $ \rm FWHM = 130 \pm 5~ km ~s^{-1}$ (corrected for instrumental broadening).   The width is narrower than the BCG's expected velocity dispersion  $\sim 250-300~\rm km ~s^{-1}$.  Molecular gas moving with a nearly isotropic velocity pattern cannot be supported against collapse at such low speeds.  The gas may be supported instead by rotation in a disk projected nearly face-on.  If so, the observed velocity width represents gas speeds out of the plane (Sec. 4.4).    

\subsection{Central Molecular Gas and Star Formation}
\label{sec: SF}
R-band and Far UV images  taken with the Hubble Space Telescope (O'Dea et al. 2010) are presented in Fig. 2.
The R-band image shows the BCG in relation to its molecular gas, hot atmosphere, and other neighboring galaxies.  The box superposed on the image 
indicates the scale of the UV and CO (3-2) images presented in Fig. 2.  A Chandra X-ray image with a similar
box superposed is shown in Fig. 2.  
Most of the molecular gas lies within the BCG's inner two arcsec diameter (8 kpc).  
The UV continuum emission is emerging from the sites of star formation proceeding at a rate upward of $100-180~\rm M_\odot ~yr^{-1}$ (M06, Egami et al. 2006, Donahue et al. 2011, ).  No bright UV or X-ray point source is associated with the AGN.  The CO (3-2) gas coincident with the UV emission is presumably fueling star formation.
The CO and UV emission are straddled by two bright and presumably rapidly cooling X-ray emission regions
oriented to the NE and SW of the nucleus.   No CO emission is detected toward the most rapidly cooling gas. Two X-ray cavities are located a few arcsec to the NW and SE of the CO (3-2) emission.  

Roughly half of the CO (3-2) flux is emerging from the inner half arcsec radius of the BCG and is unresolved.  Assuming half of the
central molecular gas mass and star formation lie within the same region, we find the surface densities of star formation and molecular gas to be $\rm log~ \Sigma_{\rm SFR} = 0.87~\rm M_\odot~yr^{-1} ~kpc^{-2}$ and 
$\rm log ~\mu_{\rm CO} = 3.2~\rm M_\odot ~pc^{-2}$, respectively.  Based on these values, the BCG lies with normal, circumnuclear starburst galaxies on the Schmidt-Kennicutt relation (Kennicutt 1998).

\subsection{Velocity Field of the Molecular Gas}

We present a grid of CO (3-2) emission spectra corresponding to the grid projected onto the CO (3-2)
image in Fig. 3.  The mean line of sight velocities measured with gaussian profile fits are indicated in each grid box.  The size of each grid box corresponds approximately to the FWHM of the synthesized beam.   Velocity differences
of a few to a few tens of $\rm km~s^{-1}$  are observed across the central structure.  No clear evidence for rotation
is observed.  If the CO (3-2) structure is a rotating disk, the small velocity gradients and narrow line width are consistent with 
it being nearly face-on.

Fig. 4 is similar to Fig. 3, but with a coarser grid intended to increase the signal in the outer region of the
central structure.   The mean line of sight velocities of the emission features are indicated  where significant CO (3-2) is detected in emission. The  tongue of gas located 1.5 arcsec to the north has a broad line profile with velocities of $-15 \rm ~ to -60~~\rm km~s^{-1}$. Likewise, the tongue extending to the west is traveling at a velocity of $ \rm -70~\rm km~s^{-1}$ with respect to the bulk of the gas.  The gas in the E-SE (bottom left) grid boxes has velocities similar in magnitude but opposite in sign (redshifted) with respect to the central emission.  The N and W tongues of blueshifted gas are oriented roughly toward the NW X-ray cavity.  The
redshifted gas to the SE is oriented roughly toward the SE X-ray cavity.   The tongues of gas appear to
be dynamically decoupled from the central structure.  Below we relate this gas to more extended molecular gas seen in CO (1-0).  We interpret these features as gas lifted out of the central region by the AGN.

We examine the molecular gas velocities on larger scales using the
grid of spectra in CO (1-0) presented in Fig. 5, which matches the resolution of the telescope configuration.   
The sky grid corresponds to spectra shown in the right panel.  The contours represent CO
emission and their colors correspond to the color coded velocity stripes superposed on the spectra. A two dimensional
gaussian profile has been fitted to and subtracted from each channel in order to remove the central emission from
the contour map. 

The CO (1-0) map reveals tongues of emission projecting roughly 10 kpc to the N-NW  and SE of the nucleus.
Their orientations are similar to the smaller, tongue-like features seen in the CO (3-2) image.
 Molecular gas traveling at $+480~ \rm km
~s^{-1}$ in the eastern box is redshifted with respect to the systemic velocity.
A narrower, blueshifted gas velocity component is seen in the N and NW boxes
extending to velocities of $-200~ \rm km~ s^{-1}$. The redshifted gas contour 
north of the nucleus is significant only at the $2-3\sigma$ level.  
Present as a small bump in the nuclear spectrum at a velocity of $300 ~\rm
km~s^{-1}$,  it is of marginal significance and will not
be discussed further.   

To summarize:  blueshifted gas lies
exclusively to the N-NW, while redshifted gas lies primarily to the
E-SE.   No significant high velocity gas
is observed in the NE, SW, and W grid boxes, nor is it observed toward the nucleus.
This pattern is consistent with a broad, bipolar flow of molecular gas, which we
discuss in greater detail in Sec. 4.3.   While this interpretation seems to
be most straightforward, it is not unique.
The gas may in principle have accreted with some net angular momentum that
placed it on nearly circular rather than radial orbits, so that the gas is in
nearly ordered motion about the BCG.  

The flux integrals under the redshifted and blueshifted
emission profile wings are $0.4 \pm 0.2$ and $0.27 \pm 0.09~ \rm Jy~km
~s^{-1}$, respectively, giving a total flux integral of $0.7 \pm 0.2~
~\rm Jy~km ~s^{-1}$. They correspond to a molecular hydrogen mass of
$1.0\pm0.3\times 10^{10}~ \rm M_\odot$.  The accuracy of the flux integrals are
sensitive to the continuum, particularly in the redshifted emission wing.
A slightly higher mass is observed in the redshifted
component compared to the blueshifted component, indicating an asymmetric flow. 

\section{Discussion}
\subsection{Bipolar Outflow or Inflow of Molecular Gas?}

Because the high speed molecular gas is observed in emission, the ALMA
observations alone cannot discriminate with certainty
between inflow and outflow.  An inflow of molecular gas from the cooling flow is the most natural explanation.   
However, if the gas cooled in a steady accretion flow 
the highest  speed gas would be observed toward the nucleus and the lowest speed gas at larger radii  (Lim et al. 2008).   We
do not observe this.  The highest speed gas shown in Fig. 5 avoids the
nucleus.  Assuming the CO (3-2) and CO (1-0) track the same dynamics,
a trend of increasing line-of-sight gas velocities with increasing radius is observed. 
The gas to the N-NW is blueshifted from velocities of a few tens of $\rm km~s^{-1}$
at radii of $\sim 3 ~\rm kpc$ to $\sim 250~ \rm km~s^{-1}$ at a radius of $\sim 10~\rm kpc$.
Likewise, the gas to the E-SE is redshifted with velocities of $\sim 40-60~ \rm km~s^{-1}$
at $\sim 3 ~\rm kpc$, increasing to $\sim 480~ \rm km~s^{-1}$ 
at $\sim 10~\rm kpc$.  The yellow wing at  $+250~ \rm km~s^{-1}$ 
 in the nuclear spectrum becomes stronger and redshifted to
higher velocities approaching the eastern grid box.   Taken together, the velocity patterns in both
CO (3-2) and CO (1-0) emission are consistent with an accelerating, bipolar outflow of molecular gas.

\subsection{Driving a Molecular Outflow by Radiation Pressure or Supernovae}

Molecular outflows are common in ULIRGs, QSOs, and starburst galaxies (reviewed by Veilleux 2005 and Fabian 2013).  Driving mechanisms include radiation pressure on dust and mechanical winds driven by supernovae.  Radiation from hot stars and AGN will  drive out gas when $dM/dt \times v_{\rm CO} \lae L_{\rm UV}/c$.  The left hand side is the product of the outflow rate and the gas velocity.  The right hand side is the sum of the UV luminosity from the AGN and stars divided by the speed of light.   The far UV image shown in Fig. 2 reveals no UV point source
associated with the AGN.  Stars are producing all of the UV flux.  
Based on the far UV flux from O'Dea et al. (2010),  $L_{\rm FUV}=1.85 \times 10^{43}~\rm erg~s^{-1}$, and assuming an outflow rate of $200 ~\rm M_\odot ~yr^{-1}$, radiation pressure is too feeble to accelerate it by more than 3 orders of magnitude. 

The power input by core collapse supernovae, $4\times 10^{43}
\rm ~ erg~s^{-1}$ (M06), is comparable to the kinetic power of the outflow, $ E_{\rm K} \simeq 10^{58} \rm erg$, $t_{\rm out} \simeq 3\times 10^7 ~\rm yr$, $P_{\rm K} \sim 10^{43}~ \rm erg~s^{-1}$.  However, for supernovae to accelerate the flow,  most of their mechanical energy must be coupled to molecular gas. Little may be thermalized.  Furthermore, their spherical blast patterns would have difficulty driving a bipolar flow.  The additional work that must be done against gravity and the surrounding pressure of the hot atmosphere would hinder a sustained and substantial bipolar flow.  While supernovae are unlikely to accelerate the flow appreciably,  they may puff-up the disk and increase the cross section between the molecular gas and the AGN,  discussed further in Sec. 4.4.

\subsection{A Radio-AGN Driving a Molecular Outflow}

The mechanical power of the jet estimated from the X-ray cavities, $P_{\rm cav} \simeq 10^{45}\rm ~erg~s^{-1}$, is by far the most potent power source.  Although the jet momentum is insufficient to lift the gas, the kinetic energy of the cold flows is only $\sim 1\%$ of the total energy output of the AGN and could easily be supplied by it.   The gas appears to be moving along and behind the rising bubbles, providing circumstantial evidence connecting them to the molecular flow.  Moreover, the molecular flow speeds are consistent with the buoyancy speeds of the cavities.  Bubbles  rise at a substantial fraction of the atmosphere's sound speed  (Churazov et al. 2001), which for Abell 1835 is $\sim 1000 ~\rm km ~s^{-1}$.  
 
 Despite ample AGN power, the bubbles must, in the end, couple to the gas and lift it out of the galaxy. 
 By Archimedes principle, they would be unable to lift more molecular gas than hot gas they displace, which is $\sim 3\times 10^{10}~\rm M_\odot$.   Rising X-ray bubbles not only displace gas, they draw metal-enriched X-ray plasma out from cluster centers at rates
of tens to hundreds of solar masses per year  (Simionescu et al. 2008, Werner et al. 2010, Kirkpatrick et al. 2009, 2011).  Abell 1835's AGN is lifting $\sim 4 \times 10^{10}~\rm M_{\odot}$ of hot gas out along the bubble axis (Kirkpatrick in preparation),
a value that is a few times larger than the mass of the molecular outflow and close to our estimate of the amount
of gas displaced by the bubbles.  Nevertheless, these estimates are uncomfortably close to the outflowing
molecular gas mass, and would imply surprisingly efficient coupling between the radio jets and bubbles
and both the hot, $\sim 4 \times 10^7 \rm ~K$, tenuous, $0.1\rm ~cm^{-3}$, volume-filling plasma and the $\sim 30$ K molecular gas.  

Understanding how the molecular gas couples to the bubbles is a challenging problem.  Both the hot and cold flows may be related. Hydrodynamical simulations have shown that ram pressure associated with high Eddington ratio jets are able to couple efficiently to both the cold and warm phases of the interstellar medium of galaxies and drive some of it out (e.g., Wagner, Bicknell, \& Umemura 2012).  Whether the molecular clouds in Abell 1835 are being accelerated by jets or are being lifted in the updraft of the X-ray cavities  (e.g., Pope et al. 2010) is unclear.  Observation suggest the latter.  Molecular gas is more readily coupled to the hot gas if the density contrast is low.  In section 4.4, the density in the molecular disk is estimated to be $\sim 1000$ times that of the hot phase.  However, for a turbulent velocity of $100\rm\ km\ s^{-1}$, if the dynamical pressure of the molecular gas matched the pressure of the hot gas, it would only be ~50 times as dense.  The high level of turbulence maintained by rapid ongoing star formation may help to explain how the tenuous hot gas is able to lift molecular gas.
 
The bubbles would be able to lift the mass more easily if the
molecular hydrogen cooled out of hotter gas rising up in the bubbles' wakes, similar to those discussed by Revaz et al. (2008).  The mean plasma density and temperature
in the central 20 kpc of the cluster is  $0.1~\rm cm^{-3}$  and $T=2.5$ keV, respectively.
The cooling time of this gas, assuming solar metal abundance, is 0.22 Gyr, which is longer than the time it  would take to displace the gas to the observed projected distance of about 10 kpc.   However, any lower entropy gas at, say, 1 keV in local pressure 
equilibrium would have a density of $0.25~\rm cm^{-3}$,
and its cooling time would be only $2\times 10^7\rm yr$.  This timescale is comparable to the rise times of the bubbles 
and molecular gas.  Therefore,
the plasma lifted out of the center with temperatures of 1 keV and below would have time to cool behind the cavities.  This plasma would have densities only a few times larger than the ambient plasma.  However,  it would be several orders of magnitude less dense than the molecular gas itself,  making it much easier to lift and accelerate to the speeds observed.  This mechanism
may help to explain the dearth of gas cooling below 1 keV in Abell 1835 and other clusters (Peterson et al. 2003). 
 
 The outflow rate is a poorly defined quantity. We estimate it by dividing the mass of the outflowing molecular gas by either the time 
 for the bubbles to rise by buoyancy to their current projected distances or by the time required for the molecular gas to
 reach its current projected distance from the nucleus.  A range of velocities and distances are observed, with timescales
 between $3-5\times 10^7~\rm yr$.  They imply an outflow rate of $\sim 200-300 ~\rm M_\odot ~yr^{-1}$. 

This rate is comparable to the BCG's mean star formation rate.  The center of the galaxy could be swept of its molecular gas in only a few hundred million years, starving the black hole and starburst of needed fuel.  However, the fate of most of the gas is unclear.  The one dimensional outflow speeds are somewhat larger than the circular speed of the stars.   If the molecular gas is flowing ballistically, most of it should return unless it evaporates into the hot atmosphere.  
If the molecular gas is coupled to the rising bubbles or continues to be
accelerated by the AGN, it could travel further.   However, if the molecular gas formed behind the bubbles in a cooling wake, it is unlikely to evaporate into the hot medium.  Most may return in a circulation flow or  "fountain" of molecular gas, perhaps similar to that inferred in  NGC 1275 (Lim et al. 2008, Salome et al. 2006, 2011).  The cycling of molecular gas in BCGs and normal elliptical galaxies could have a significant impact on their star formation and AGN histories.

\subsection{Dynamics of the Central Molecular Gas}

The dynamical state and high average density of the molecular gas in
the central kiloparsec of the BCG have significant implications for
this system.  The lack of evidence for rotation in the molecular gas
implies that, if the gas is rotationally supported, its rotation axis
must be very close to our line of sight.  At the same time, the full
velocity width at half maximum for the molecular gas in this region is
$130~\rm km~s^{-1}$, corresponding to a line of sight velocity dispersion 
of $\sigma_{\rm los}\simeq 55\rm ~km ~s^{-1}$ and a one-dimensional
turbulent velocity $v_T=\sigma_{\rm los}$.  This suggests the gas lies
in a face-on disk.



This high turbulent velocity is consistent with the disk being
marginally gravitationally unstable; the Toomre Q parameter is 
\begin{eqnarray}
Q&=&{v_cv_T\over \pi G r\Sigma_g(r)}\approx0.43
\left({v_c\over 400~\rm km~s^{-1}}\right)
\left({v_T\over 55~\rm km~s^{-1}}\right)\nonumber\\
&&\left({R_e\over 2 ~\rm kpc}\right)^{-1}
\left({\Sigma_g(R_e)\over 0.4 ~\rm g~cm^{-2}}\right)^{-1},
\end{eqnarray}
where we have scaled to the radius $R_e$ enclosing half the CO (3-2)
flux, which we assume to enclose half the mass. We have also scaled the circular velocity at that radius to
$400~\rm km~s^{-1}$. The circular velocity would have to exceed $940~\rm km~s^{-1}$ to
stabilize the gas disk. We noted in Section~\ref{sec: SF} that the BCG lies
with starburst galaxies on the Schmidt-Kennicutt relation; galaxies on
that relation have $Q\approx1$. 

We note that for a gas to dust mass ratio of 100, the surface density
($\Sigma_g=1600~\rm M_\odot\pc^{-2}$) corresponds to an $A_v\approx100$. 

The weight per unit area under self-gravity, or dynamical pressure of
this gas is 
\be
p_{\rm dyn}={\pi\over2} G\Sigma_g^2\approx1.7\times10^{-8}{\rm dyne}~\rm cm^{-2}.
\ee
This is substantially higher than the thermal pressure of the hot gas.

Being marginally gravitationally stable implies that the disk scale
height $H=(v_T/v_c)r$, or about $275~\rm pc$ at $R_e=2~\rm kpc$. The mean
density at that radius is then
$\bar\rho_c=\Sigma_g/(2H)\approx2.4\times 10^{-22}~\rm g~ cm^{-3}$, or
$n_H\approx100$. This mean density is somewhat higher than the mean
density in massive Milky Way GMCs. 
The Toomre mass of molecular clouds is then $M_T = H^2 \Sigma_g \simeq 1.4\times10^8~\rm M_\odot$.
The turbulent pressure of the cold
gas is  \be \label{eqn: turbulent pressure} p_{\rm turb}=\bar\rho_c
v_T^2\approx 0.7\times 10^{-8}{~\rm dyne}~\rm cm^{-2}, \ee i.e., the turbulent
motions provide enough pressure to support the disk in a marginally
stable state.

Turbulence is believed to decay on a dynamical time. Maintaining the
turbulence in A1835 would then require a turbulent power of
\bd
P_{\rm turb}={3 M_g v_T^2 \over 2 R_e / v_c} \simeq  3 \times 10^{43}~\rm erg~s^{-1}.
\ed
This is similar to the total luminosity supplied by supernovae, if the star formation rate is $\sim 200~\rm M_\odot ~yr^{-1}$, $L_{Sne}=6\times10^{43}(\dot M_*/200~\rm M_\odot~yr^{-1})~erg~s^{-1}$, if the supernovae are well coupled to the molecular gas, and if they do not radiate more than $\sim 50\%$ of their energy away.   We observe that $Q\approx1$, so we expect large scale features such as spiral arms or bars in the disk to develop. These features will transport angular momentum efficiently inward, down to sub-parsec scales, e.g. Hopkins \& Quataert (2010, 2011).


\subsection{Origin of the Molecular Gas}

Molecular gas associated with starburst galaxies, ULIRGS, and QSOs is usually attributed to wet mergers.  
The center of a rich cluster with a large velocity dispersion is a difficult place for this to happen.
Ram pressure experienced by a plunging, gas-rich donor galaxy will strip most of its atomic gas and much
of its molecular gas before 
it reaches the BCG (Combes 2004, Roediger \& Br\"uggen 2007, Kirkpatrick et al. 2009, Ruszkowski et al. 2012).
Molecular gas, being dense and centrally concentrated, is tightly bound and more resilient to stripping than atomic gas.
Therefore, short of a direct collision, a plunging galaxy should retain much of its molecular gas (Young et al. 2011).   
Furthermore, the BCG's molecular gas mass exceeds by large factors that of most galaxies in clusters at its epoch.  
The likelihood that such a galaxy, if present, would hit the BCG directly and deposit its molecular gas 
at the low speeds observed seems remote.

The molecular gas in Abell 1835 probably cooled from the hot atmosphere and settled into the BCG.
Hot atmospheres whose central cooling time falls below $\sim 10^9~\rm yr$
frequently harbor $10^9 -10^{10}~\rm M_\odot$ of molecular gas (Edge 2001, Salome \& Combes 2003).  BCGs in Coma-like
clusters with long central cooling times do not.   Abell 1835 is the
extreme of this class.  Its cooling rate of $\lae 140 ~\rm M_\odot ~yr^{-1}$ (Sanders et al. 2010)  would supply 
the molecular gas in a few hundred Myr, comparable to the age of the starburst (M06).     Molecular gas
condensing from a hot atmosphere will achieve lower velocities and lower specific angular momentum
than gas arriving by merger (Lim et al. 2008).  Sensitive ALMA observations in future may detect molecular gas
cooling and flowing in from the hot atmosphere and returning in a circulation flow.

\section{Conclusions}

We have shown that the BCG in Abell 1835 contains roughly $5\times 10^{10}~\rm M_{\odot}$ of molecular
gas, most of which is associated with stars forming at $\gae 100 ~\rm M_{\odot}~yr^{-1}$, possibly in a thick, face-on disk. 
We discovered a $\sim 10^{10}~\rm M_{\odot}$ bipolar molecular outflow traveling between $-250~\rm and~ +480 ~\rm km~s^{-1}$  that is being accelerated most likely by mechanical energy associated with rising X-ray bubbles. 
Whether the bubbles
accelerated the molecular clouds directly, or whether the molecular clouds cooled
out of the hot plasma in the updraft behind the bubbles isn't clear.  Both mechanisms have problems.
The mass of hot plasma being displaced and dragged out by the bubbles, $3-4\times 10^{10}~\rm M_\odot$, is uncomfortably similar to the $10^{10}~\rm M_\odot$ of molecular gas they apparently lifted.  Direct uplift would require strong coupling between dense, molecular clouds and the tenuous atmosphere, which may be difficult to achieve.  
 The problem would be lessened if the molecular gas cooled in updraft behind the bubbles, or if the outflow 
 mass were lower than we have estimated. For example, the $\rm X_{\rm CO}$ parameter may be lower than the
 Galactic value we have assumed.



Our result has broader implications.  Molecular gas abundance is a sharply declining function of a galaxy's stellar mass.  Above 
$3\times 10^{10}~M_\odot$ most are elliptical galaxies.  Of these, only $\sim 22\%$ contain molecular gas, and only at levels between $10^7-10^9~\rm M_\odot$ (Young et al. 2011).   On the other hand, radio power is a steeply increasing function of stellar mass
(Best et al. 2005, Best \& Heckman 2012).  Their detection fraction rises from $0.01\%$ at $3\times 10^{10}~M_\odot$
to upward of $30\%$ at $5\times 10^{11}~M_\odot$ (Best et al. 2005).  Therefore, molecular gas mass must also be
a declining function of radio power.  While a number of environmental factors may be contributing to this decline (Young et al. 2011), the radio source itself may play a role, albeit a complicated one.  Radio synchrotron power represents only a small fraction of a 
radio AGN's total mechanical power (Birzan et al. 2008).  Therefore, relatively low
power radio synchrotron sources can be mechanically potent. 
Mechanical heating of hot atmospheres in elliptical galaxies by radio mode feedback is likely to be the primary mechanism maintaining ``red and dead" elliptical galaxies (e.g., Bower et al. 2006, Croton et al. 2006).  However, if AGN are fed by cold molecular clouds, 
a feedback loop may be difficult to sustain unless the radio jets are also regulating the rate of cold accretion.
The relatively efficient coupling between the molecular gas and radio bubbles implied by the Abell 1835 observation
suggests that  radio mode feedback may also regulate the amount of molecular gas 
reaching the nuclei of galaxies.    

BRM thanks Tom Jones and Christine Jones for helpful comments.
HRR and BRM acknowledge generous financial support from the Canadian
Space Agency Space Science Enhancement Program.  BRM, RAM, HRR, and ANV
acknowledge support from the Natural Sciences and Engineering Research
Council of Canada. ACE acknowledges support from STFC grant ST/I001573/1
 PEJN is supported by NASA grant NAS8-03060.
We thank the ALMA scientific support staff members
Adam Leroy and St\'ephane Leon.  The paper makes use of the following
ALMA data: ADS/JAO.ALMA\#2011.0.00374.S. ALMA is a partnership of ESO
(representing its member states), NSF (USA) and NINS (Japan), together
with NRC (Canada) and NSC and ASIAA (Taiwan), in cooperation with the
Republic of Chile. The Joint ALMA Observatory is operated by ESO,
AUI/NRAO and NAOJ.  The National Radio Astronomy Observatory is a
facility of the National Science Foundation operated under cooperative
agreement by Associated Universities, Inc.  This paper is dedicated to Jim Pisano,
who helped make ALMA the marvelous facility it is.

\appendix
\section{The CO to H2 Conversion Factor}

CO traces traces  molecular hydrogen which, lacking a permanent electric dipole moment, radiates inefficiently.  The value of the CO to molecular gas conversion factor, commonly referred to as $\rm X_{\rm CO}$, is the prime uncertainty in our mass estimates.  Absent an alternative, most investigators adopt the value for the Milky Way Galaxy and other local disk galaxies, where the CO (1-0) emission feature is usually optically thick.  However, the true value depends on environmental factors, such as the gas phase metal abundance, which may depart from the average Galactic value.  A lower gas phase metal abundance gives a higher mass ratio of hydrogen to CO.  Therefore, applying the Galactic $\rm X_{\rm CO}$ to low metal abundance gas would underestimate of the total molecular gas mass.  On the other hand, if the molecular gas optically thin or nearly so, as it may be in turbulent flows and massive starburst galaxies, the Galactic $\rm X_{\rm CO}$ would over estimate the molecular gas mass.  Other factors that affect  $\rm X_{\rm CO}$ including, the temperature, density, and dynamics of the gas, which in most situations are poorly understood (Bolatto et al. 2013).   

The metallicity of the cooling X-ray plasma in Abell 1835 lies between 0.5-0.8 times the Solar metallicity within 20 kpc of the nucleus.  This alone would indicate that adopting the Galactic $\rm X_{\rm CO}$  as we have done should provide a reasonable if not a conservative underestimate of the molecular gas mass.  However, Abell 1835 is a starburst galaxy. There are indications that $\rm X_{\rm CO}$ in starburst galaxies may be depressed below the Galactic value.  The central gas density, $\sim 2000~ M_\odot \rm ~pc^{-2}$, lies midway between normal spirals and starbursts.  The gas density of the outflow, away from the bulk of star formation, has a surface density of
$\sim 100 ~M_\odot ~ \rm pc^{-2}$, which is comparable to normal spiral galaxies and to the Milky Way (Bolatto et al. 2013).  It is therefore possible that the $\rm X_{\rm CO}$ value for the molecular gas located near the nucleus may be suppressed by a small factor with respect to the molecular gas in the outflow.  On the other hand,  indications are that $\rm X_{\rm CO}$  may be suppressed in turbulent winds, where the molecular gas becomes optically thin (Bolatto 2013).  Abell 1835's outflow velocity is lower than those in quasars (e.g., Maiolino et al. 2012, Feruglio et al. 2010).  Taken together, we have no reason to expect $\rm X_{\rm CO}$ to depart significantly from the Galactic value in this system.  Nevertheless, should $\rm X_{\rm CO}$ lie a factor of several below the Galactic value, the flow would still exceed $10^9~ \rm M_\odot$. This would not qualitatively alter our result.

\centerline{References}

Best, P.~N., Kauffmann, G., Heckman, T.~M., et al.\ 2005, \mnras, 362, 25 \\

Best, P.~N., \& Heckman, T.~M.\ 2012, \mnras, 421, 1569 \\

B{\^i}rzan, L., McNamara, B.~R., Nulsen, P.~E.~J., Carilli, C.~L., \& Wise, M.~W.\ 2008, \apj, 686, 859 \\

Bower, R.~G., Benson, A.~J., Malbon, et al.\ 2006, \mnras, 370, 645 \\

Bolatto, A.~D., Wolfire, M., \& Leroy, A.~K.\ 2013, arXiv:1301.3498 \\

Cavagnolo, K.~W., Donahue, M., Voit, G.~M., \& Sun, M.\ 2008, \apjl, 683, L107 \\

Churazov, E., Br{\"u}ggen, M., Kaiser, C.~R., B{\"o}hringer, H.,  \& Forman, W.\ 2001, \apj, 554, 261 \\

Combes, F.\ 2004, Recycling Intergalactic and Interstellar Matter, 217, 440 \\

Crawford, C.~S., Allen, S.~W., Ebeling, H., Edge, A.~C., \& Fabian, A.~C.\ 1999, \mnras, 306, 857 \\

Croton, D.J. et al. 2006,  MNRAS, 365, 11\\

Donahue, M., de Messi{\`e}res, G.~E., O'Connell, R.~W., et al.\ 2011, \apj, 732, 40 \\

Edge, A. C. 2001, MNRAS, 328, 762\\

Egami, E., Misselt,  K.~A., Rieke, G.~H., et al.\ 2006, \apj, 647, 922 \\

Fabian, A.~C.\ 1994, \araa, 32, 277\\

Fabian, A.C. 2013, ARA\&A, 50, 455\\

Feruglio, C., Maiolino, R., Piconcelli, E., et al.\ 2010, \aap, 518, L155 \\

Gaspari, M., Ruszkowski, M., \& Sharma, P.\ 2012, \apj, 746, 94 \\

Gaspari, M., Ruszkowski, M., \& Oh, S.~P. 2013, \mnras, 432, 3401\\

Guo, F., \& Mathews, W.~G.\ 2013, arXiv:1305.2958\\
 
Heckman, T.~M.\ 1981, \apjl, 250, L59\\

Hopkins, P.~F. \& Quataert, E. \ 2010, \mnras, 407, 1529\\

Hopkins, P.~F. \& Quataert, E. \ 2011, \mnras, 415, 1027\\

Hu, E.~M., Cowie, L.~L., \& Wang, Z.\ 1985, \apjs, 59, 447 \\

 Kennicutt, R.~C., Jr.\ 1998, \apj, 498, 541 \\

Kirkpatrick, C.~C., McNamara, B.~R., \& Cavagnolo, K.~W.\ 2011, \apjl, 731, L23 \\

Kirkpatrick, C.~C., Gitti, M., Cavagnolo, K.~W., et al.\ 2009, \apjl, 707, L69 \\

 Lim, J., Ao, Y., \& Dinh-V-Trung 2008, \apj, 672, 252 \\

Maiolino, R., Gallerani, S., Neri, R., et al.\ 2012, \mnras, 425, L66 \\

McDonald, M., Bayliss, M., Benson, B.~A., et al.\ 2012, \nat, 488, 349 \\

 McNamara, B.~R., \& Nulsen, P.~E.~J.\ 2007, \araa, 45, 117 \\

McNamara, B.R.,  Nulsen, P.E.J., 2012, NJP, 14, 055023 \\

McNamara, B.~R., Rafferty, D.~A., B{\^i}rzan, L., et al.\ 2006, \apj, 648, 164 \\

Morganti, R., Tadhunter, C.~N., \& Oosterloo, T.~A.\ 2005, \aap, 444, L9 \\

Morganti, R., et al.  2013, Science, in press\\

Nesvadba, N.~P.~H., Lehnert, M.~D., Eisenhauer, F., et al.\ 2006, \apj, 650, 693 \\

 O'Dea, C.~P., Baum, S.~A., Privon, G., et al.\ 2008, \apj, 681, 1035 \\

O'Dea, K.~P., Quillen, A.~C., O'Dea, C.~P., et al.\ 2010, \apj, 719, 1619 \\

O'Neill, S.~M., \& Jones, T.~W.\ 2010, \apj, 710, 180 \\

Peterson, J.~R., \& Fabian, A.~C.\ 2006, \physrep, 427, 1 \\

Peterson, J.~R., Kahn, S.~M., Paerels, F.~B.~S., et al.\ 2003, \apj, 590, 207 \\

Pizzolato, F. \& Soker, N. 2010, \mnras, 408, 961\\

Pope, E.~C.~D., Babul, A., Pavlovski, G., Bower, R.~G., \& Dotter, A.\ 2010, \mnras, 406, 2023\\
 
Rafferty, D.~A., McNamara, B.~R., \& Nulsen, P.~E.~J.\ 2008, \apj, 687, 899 \\

Revaz, Y., Combes, F., \& Salom{\'e}, P.\ 2008, \aap, 477, L33 \\

 Roediger, E., \& Br{\"u}ggen, M.\ 2007, \mnras, 380, 1399 \\

Ruszkowski, M., Bruggen, M., Lee, D., \& Shin, M.-S.\ 2012, arXiv:1203.1343 \\

Sanders, J.~S., Fabian, A.~C., Smith, R.~K., \& Peterson, J.~R.\ 2010, \mnras, 402, L11 \\

Salome, P., \& Combes, F. 2003, A\& A, 412, 657\\

Salom{\'e}, P., Combes, F., Edge, A.~C., et al.\ 2006, \aap, 454, 437 \\

Salom{\'e}, P., Combes, F., Revaz, Y., et al.\ 2011, \aap, 531, A85 \\

Simionescu, A., Werner, N., Finoguenov, A., B{\"o}hringer, H., \& Br{\"u}ggen, M.\ 2008, \aap, 482, 97 \\

Veilleux, S., Cecil, G., \& Bland-Hawthorn, J.\ 2005, \araa, 43, 769 \\

Voit, G.~M.\ 2011, \apj, 740, 28 \\

Voit, G.~M., Cavagnolo, K.~W., Donahue, M., et al.\ 2008, \apjl, 681, L5 \\

Wagner, A.~Y., Bicknell, G.~V., \& Umemura, M.\ 2012, \apj, 757, 136\\
  
Werner, N., Simionescu, A., Million, E.~T., et al.\ 2010, \mnras, 407, 2063 \\

Wilman, R.~J., Edge, A.~C., \& Swinbank, A.~M.\ 2006, \mnras, 371, 93 \\

Young, L.~M., Bureau, M., Davis, T.~A., et al.\ 2011, \mnras, 414, 940 \\


\newpage

\begin{figure} 
    \epsscale{1.0}
    \plottwo{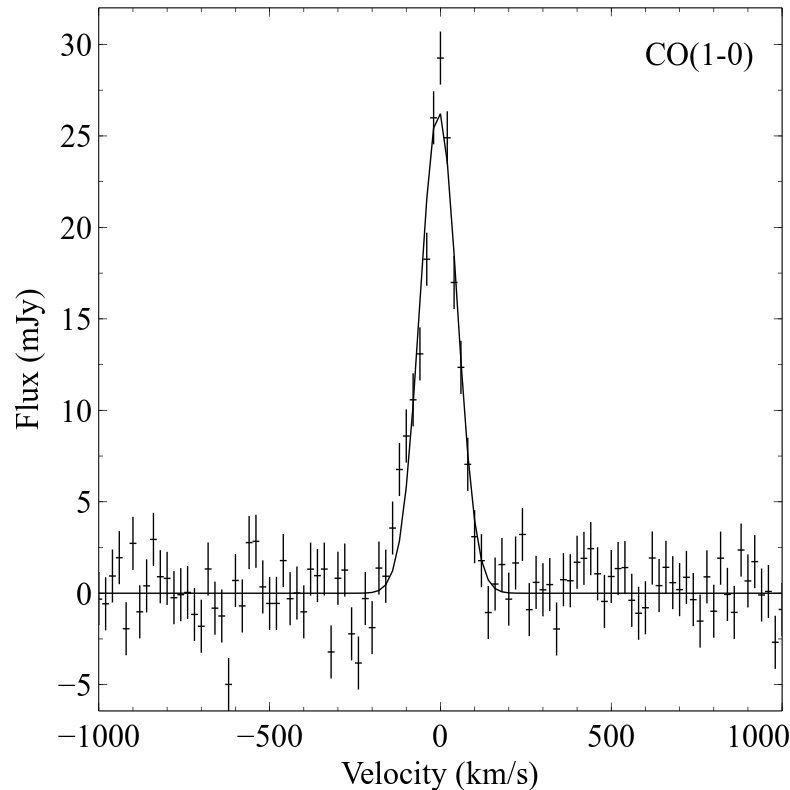}{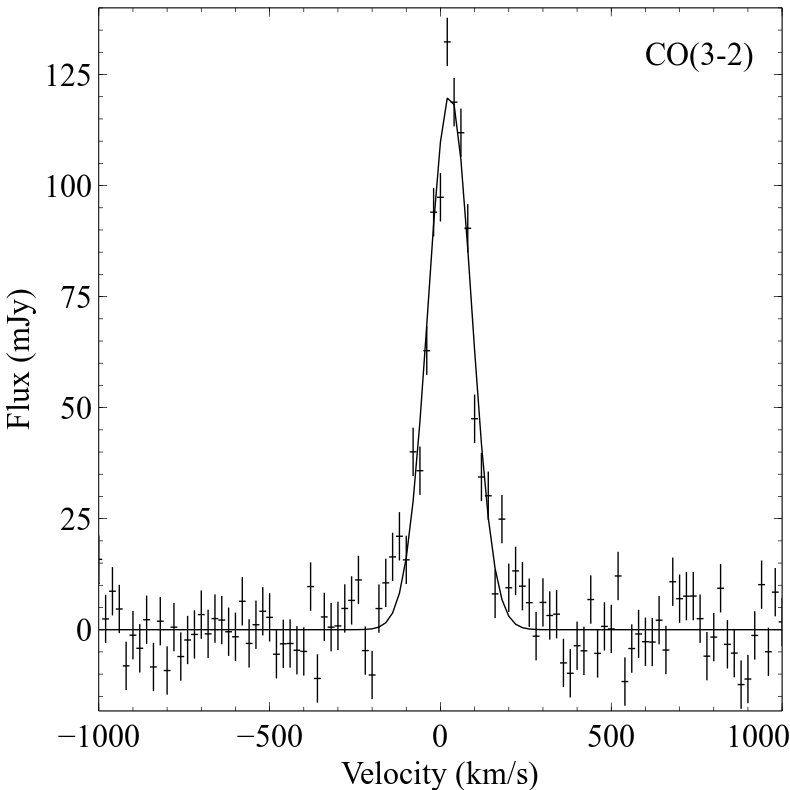}
    
  \caption{CO (1-0) and CO (3-2)  spectra on the left and right, respectively.  The spectra were extracted from regions measuring $6 \times 6$ arcsec. }
\end{figure}

\begin{figure} 
    \epsscale{1.25}
    \plotone{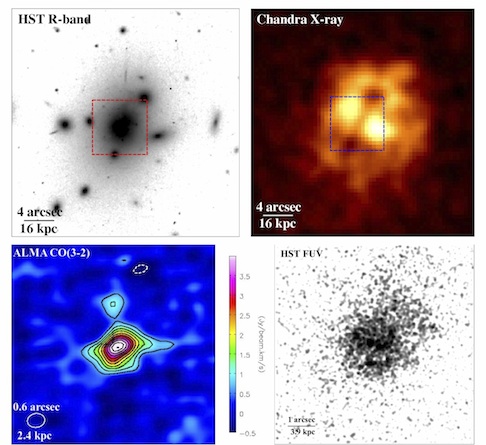}
  \caption{{\bf Upper left:} Hubble Space Telescope F702W WFPC2  image of the BCG and surrounding galaxies in Abell 1835. The red box indicates the scale of the CO (3-2) image at lower left.  The image is sensitive to both the old and young stellar populations and accompanying line emission.  {\bf Upper right:} Chandra X-ray image of the hot atmosphere surrounding the BCG. A smooth X-ray background has been subtracted.  The blue box indicates the location and scale of the CO (3-2) and UV images in the lower panels.   X-ray cavities inflated by the radio AGN (M06) are seen to the northwest and southeast near the edges of the box.  The bright regions to the northeast and southwest of center are the locations of gas with the shortest cooling time where the atmosphere is cooling rapidly. {\bf Lower left:} CO (3-2) image.  The oval at lower left indicates the beam size, shape, and scale in arcseconds and kiloparsecs.  The contours represent $-3,+3,+6,+9...\sigma$.  {\bf Lower right:} Far ultraviolet continuum image through filter F165LP ACS Solar Blind Channel.  Note the absence of a nuclear point source associated with an AGN. Essentially all of the continuum is from the young stars.}
\end{figure}

\newpage
\begin{figure} 
    \epsscale{1.25}
  \plotone{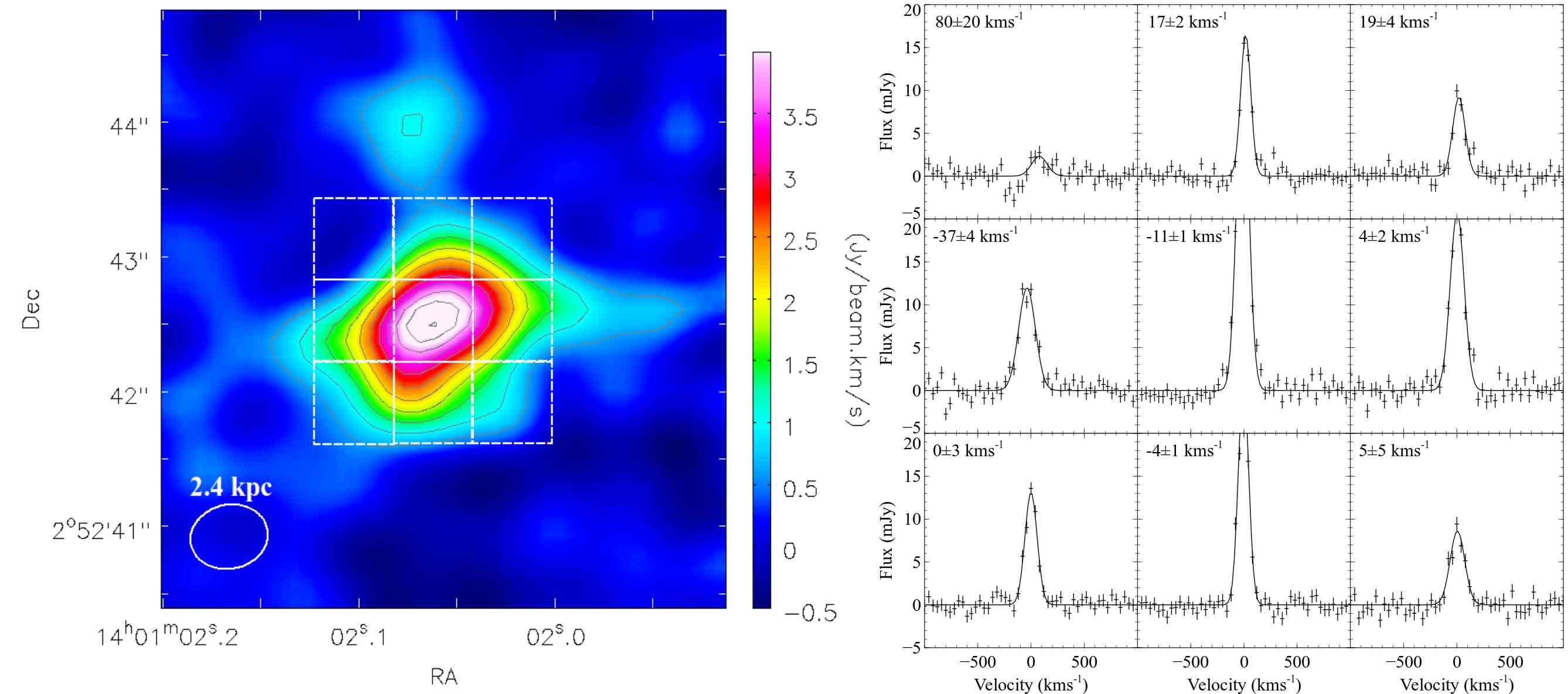}
  \caption{CO (3-2) images with a grid of apertures corresponding to the spectra shown in the right panel. The extractions are 0.6 arcsec on a side, corresponding to the resolution of the synthesized beam.  The velocity centroids and their errors are indicated in each box.  }
\end{figure}

\newpage
\pagebreak
\begin{figure} 
    \epsscale{1.25}
    \plotone{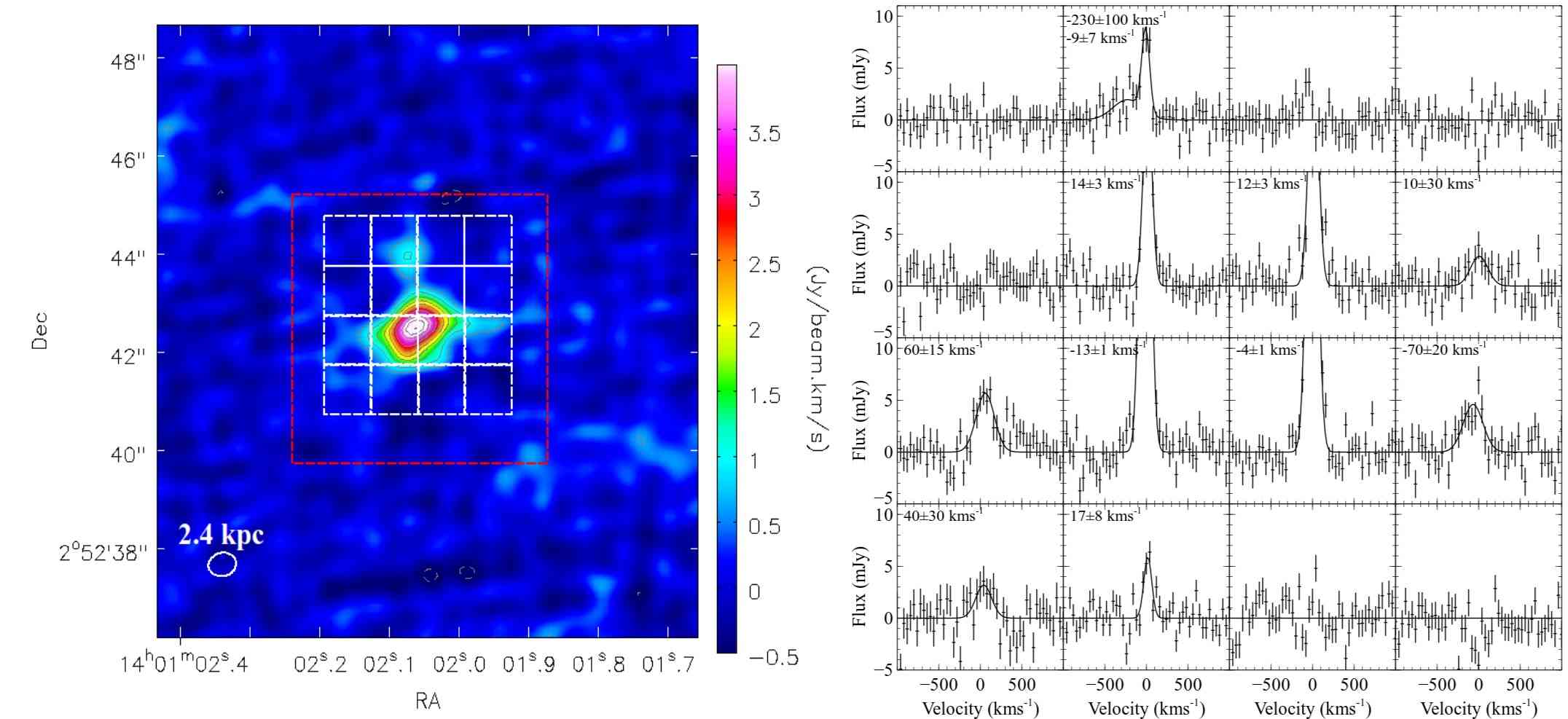}
  \caption{Similar to Fig. 3 but with 1 arcsec extraction apertures that extend into the fainter outer reaches of the
  central gas structure. The red box corresponds to the outer edge of the grid region shown in 
   the CO (1-0) map in Fig. 5}
\end{figure}

\newpage
\begin{figure} 
    \epsscale{1.25}
 \plotone{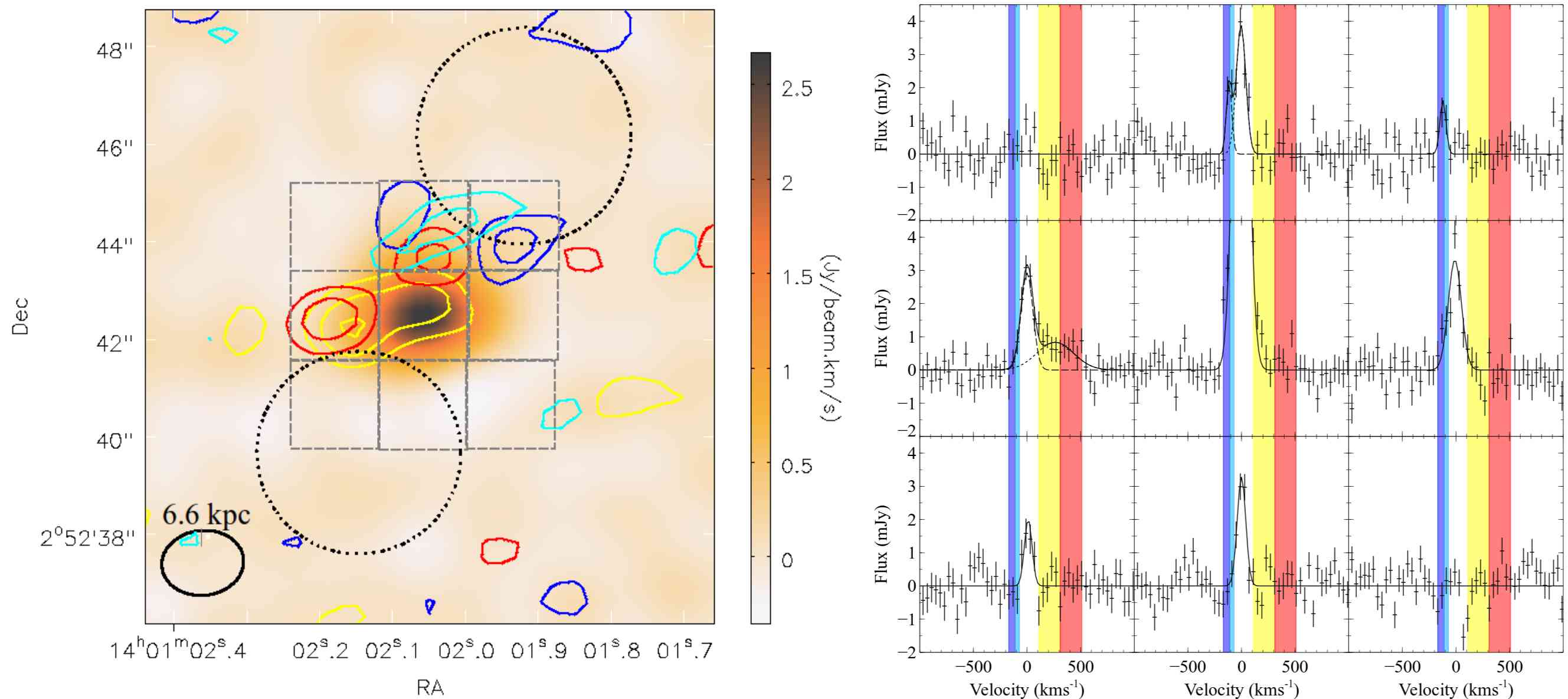}
  \caption{Left panel shows a colorscale image of CO (1-0) emission with color contours divided into separate
  velocity bins. The contours  are integrated intensity in particular velocity ranges with $+2\sigma$, 
  $3 \sigma$, $4 \sigma$.  dark blue ( $-210 \rm ~to~ -150~km~s^{-1}$, cyan ($ -150~ \rm to -110 ~ km~s^{-1}$), yellow ($70\rm ~ to~ 270~ km~s^{-1}$), red ($270~ \rm to ~470 ~km~s^{-1}$). Right panel shows spectral extractions $1.8 \times 1.8$ arcsec on a side, roughly corresponding to the CO (1-0) beam size.  The colors superposed on the spectra correspond to the velocity contours. This figure shows that the high velocity molecular gas avoids the nucleus; higher speeds are observed at larger radii, indicating outflow.  The black dotted circles show the locations of the X-ray bubbles. The molecular gas appears to be drawn up
  behind the rising bubbles.}
\end{figure}

\end{document}